\documentclass[english,aps,prd,a4paper,groupedaddress,preprintnumbers,floatfix,nofootinbib,showpacs,12pt]{revtex4}

\usepackage{graphicx}
\usepackage{enumerate}
\usepackage{amsmath}
\usepackage{amssymb}
\usepackage{amsfonts}
\usepackage{xcolor}
\usepackage{url}
\usepackage{hyperref}
\usepackage{booktabs}
\usepackage{caption}
\usepackage[utf8]{inputenc}

\usepackage{arydshln}
\usepackage{natbib}

\hypersetup{colorlinks=true,linkcolor=redLinks,citecolor=greenLinks,
urlcolor=redLinks,
pdfborder={0 0 1}}
\hypersetup{
    bookmarks=true,         % show bookmarks bar?
    unicode=false,          % non-Latin characters in Acrobat?s bookmarks
    pdftoolbar=true,        % show Acrobat?s toolbar?
    pdfmenubar=true,        % show Acrobat?s menu?
    pdffitwindow=false,     % window fit to page when opened
    pdfstartview={FitH},    % fits the width of the page to the window
    pdftitle={My title},    % title
    pdfauthor={Author},     % author
    pdfsubject={Subject},   % subject of the document
    pdfcreator={Creator},   % creator of the document
    pdfproducer={Producer}, % producer of the document
    pdfkeywords={keyword1, key2, key3}, % list of keywords
    pdfnewwindow=true,      % links in new PDF window
    colorlinks=false,       % false: boxed links; true: colored links
    linkcolor=red,          % color of internal links (change box color with linkbordercolor)
    citecolor=green,        % color of links to bibliography
    filecolor=magenta,      % color of file links
    urlcolor=cyan           % color of external links
}

\newcommand{\bi}{\begin{itemize}}
\newcommand{\ei}{\end{itemize}}

\begin{document}
\title{Future perspectives for a weak mixing angle measurement  in
  coherent elastic neutrino nucleus scattering experiments.}
\author{B. C. Ca\~nas$^{1,2}$} \email{blanca.canas@unipamplona.edu.co}
\author{E. A. Garc\'es$^1$} \email{egarces@fis.cinvestav.mx}
\author{O. G. Miranda$^1$} \email{omr@fis.cinvestav.mx}
\author{A. Parada$^3$} \email{alexander.parada00@usc.edu.co}
\affiliation{$^1$~Departamento de F\'isica, Centro de Investigaci\'on
  y de Estudios Avanzados del IPN, Apdo. Postal 14-740, 07000 Ciudad
  de M\'exico, M\'exico.}    \affiliation{$^2$~Universidad de Pamplona, Km 1, v\'ia salida
  a Bucaramanga, Campus Universitario, 543050, Pamplona, Colombia}
  \affiliation{$^3$~Universidad Santiago de
  Cali, Campus Pampalinda, Calle 5 No. 6200, 760035, Santiago de Cali,
  Colombia.} 

\begin{abstract}\noindent
After the first measurement of the coherent elastic neutrino nucleus
scattering (CENNS) by the COHERENT Collaboration, it is expected that
new experiments will confirm the observation. Such measurements will allow
to put stronger constraints or discover new physics as well as to probe the
Standard Model by measuring its parameters.  This is the case of the
weak mixing angle at low energies, which could be measured with an
increased precision in future results of CENNS experiments using, for
example, reactor antineutrinos. In this work we analyze the physics
potential of different proposals for the improvement of our current
knowledge of this observable and show that they are very promising.
\end{abstract}

\maketitle

\section{Introduction}

Neutrinos are one of the most elusive particles. With a
small cross section, its detection has been always a challenge for the
experimentalist. Despite this difficult task, neutrino physics is in a precision era
with increasingly accurate measurements~\cite{deSalas:2017kay,globalfit,Esteban:2016qun,Capozzi:2018ubv}. Among the recent progress in this
field is the detection, for the first time, of the coherent elastic
neutrino-nucleus scattering (CENNS).  This reaction was
proposed~\cite{Freedman:1973yd} just after the discovery of the weak
neutral currents~\cite{Hasert:1973cr} and recently detected by the COHERENT
collaboration~\cite{Akimov:2017ade}.  Besides the natural interest in
confirming this recent detection, there are different issues that are
of current interest in nuclear and neutrino physics.  Many new physics
scenarios can be probed, as it has been proposed in the case of Non
Standard Interactions
(NSI)~\cite{Barranco:2007tz,Barranco:2011wx,Billard:2018jnl,Denton:2018xmq},
a $Z'$ gauge
boson~\cite{Shoemaker:2017lzs,Kosmas:2017tsq,Farzan:2018gtr,Ge:2017mcq},
electromagnetic neutrino
properties~\cite{Kosmas:2015sqa,Kosmas:2015vsa} and even the case of
an sterile neutrino~\cite{Canas:2017umu,Kosmas:2017zbh,Dutta:2015nlo,Garces:2017qkm}.
Methods alternative to inverse beta decay (IBD) of reactor neutrino detection can also shed
light in the so called reactor neutrino anomaly~\cite{Mention:2011rk},
as we have pointed out in~\cite{Canas:2017umu}.

\noindent
Reactor neutrinos have a great tradition of discoveries, since the
first neutrino detection~\cite{reines:1953pu} and in the last
decades they have played an important role in establishing the three
neutrino oscillation paradigm~\cite{deSalas:2017kay}, IBD
 has been the golden channel in reactor
neutrino detection. However there are other interesting neutrino reactions that can also be
used to probe neutrino fluxes from reactors, as is the case of elastic
neutrino-electron scattering (ENES) detected for the first time in the
seventies~\cite{Reines:1976pv} and measured with increased precision
by the TEXONO~\cite{Wong:2006nx} and MUNU~\cite{Daraktchieva:2005kn}
Collaborations; and more recently of CENNS measured at the neutron spallation source by the COHERENT
Collaboration~\cite{Akimov:2017ade}.  
It is expected that in the near future improved measurements 
of ENES reaction can be provided by the GEMMA experiment~\cite{Beda:2013mta}.

\noindent
The expectation for a new measurement of the weak mixing angle in
CENNS has already been studied in the past, for example for the case
of the TEXONO~\cite{Kosmas:2015vsa} and the
CONUS~\cite{Lindner:2016wff} proposals.  Here we focus in the case
of the
CONNIE~\cite{Aguilar-Arevalo:2016qen,Aguilar-Arevalo:2016khx,Chavez-Estrada:2017gni}, 
MINER~\cite{Agnolet:2016zir}, and 
RED100~\cite{Akimov:2012aya} research programs and reanalyse the
TEXONO and CONUS case studies in order to compare them on an equal footing and to
contrast the importance of different characteristics of each
experiment. In particular, we note here how sensitivities can depend on the experiment 
detection targets due to a different protons to
neutrons proportion.

\noindent
The dependence of CENNS cross section on the weak charge $Q_W$ allows
the study of the weak mixing angle at extremely low momentum transfer,
a region where an improvement in the accuracy of this parameter is
very much needed~\cite{Erler:2004in,Garces:2011aa}, particularly in measurements
with neutrino interactions~\cite{Canas:2016vxp}. We will show that,
although the sensitivity to the weak charge is relatively small in 
CENNS, it will be possible to have competitive measurements of the
$\sin^2\theta_W$ in the low energy regime if the systematic
uncertainties are under control. We will discuss that, besides the
importance of high statistics, the proportion of protons to neutrons
in a given target will also play an important role. 

\section{ CENNS experiments with reactor antineutrinos}

Several future proposals plan to measure CENNS with increased
statistics, opening the possibility to test the Standard Model in the
ultra-low energy regime. To study the sensitivity of these proposals
to the weak mixing angle, we start by considering the CENNS cross
section, given by the following expression~\cite{Barranco:2005yy}

\begin{equation} 
\left(\frac{d\sigma}{dT}\right)_{\rm SM}^{\rm coh} = \frac{G_{F}^{2}M}{2\pi}\left[1-\frac{MT}{E_{\nu}^{2}}+\left(1-\frac{T}{E_{\nu}}\right)^{2}\right]
 [Zg^{p}_{V}F_Z(q^2)+Ng^{n}_{V}F_N(q^2)]^{2}. \label{eq:00}
\end{equation}

\noindent Here, $M$ is the mass of the nucleus, $E_{\nu}$ is the
neutrino energy, and $T$ is the nucleus recoil energy; $F_{Z,N}(q^2)$
are the nuclear form factors that are especially important at higher momentum transfer, as can be the
case of
neutrinos coming from spallation neutron sources, while for reactor
antineutrinos, they have a minimal impact and will be considered as
equal to one in this work.  The neutral current vector couplings (including
radiative corrections) are given by~\cite{Barranco:2005yy},

\begin{eqnarray}
\nonumber g_{V}^{p} &=&\rho_{\nu
  N}^{NC}\left(\frac{1}{2}-2\hat{\kappa}_{\nu
  N}\hat{s}_{Z}^{2}\right)+2\lambda^{uL}+2\lambda^{uR}+\lambda^{dL}+\lambda^{dR}\\ g_{V}^{n}
&=&-\frac{1}{2}\rho_{\nu
  N}^{NC}+\lambda^{uL}+\lambda^{uR}+2\lambda^{dL}+2\lambda^{dR}
\end{eqnarray}
\noindent where $\rho_{\nu N}^{NC}=1.0082$,
$\hat{s}_{Z}^{2}=\sin^2\theta_{W}=0.23129$, $\hat{\kappa}_{\nu
  N}=0.9972$, $\lambda^{uL}=-0.0031$, $\lambda^{dL}=-0.0025$, and
$\lambda^{dR}=2\lambda^{uR}=7.5 \times 10^{-5}$~\cite{Patrignani:2016xqp}. 

\noindent
From the previous expressions for the vector couplings, it is
straightforward to note that the dependence on the weak mixing angle
appears only on the protons coupling and, therefore, nuclei with
larger protons to neutrons proportion could be more sensitive to
this measurement.  On the negative side, we can also notice that this
contribution is small in comparison with the neutron one. Despite
this, a high statistics CENNS experiment will be sensitive to this
coupling and, therefore, the weak mixing angle can be measured with a
precision similar to the one at current measurements in this low energy
regime. Currently, most of the proposals are working with a relatively
small amount of material and considering upgrades in the near
future. In what follows, we will consider the optimistic case of the
upgraded, high statistics, detectors that are the ones that have the
possibility to make an accurate measurement.

\noindent
For estimating the number of expected events (SM) in the detector, we
use the expression,
\begin{equation}\label{n_events}
 N^{\rm SM}_{\rm events}=t\phi_{0}\frac{M_{\text{detector}}}{M}\int_{E_{\nu \rm min}}^{E_{\nu \rm max}}
 \lambda(E_{\nu})dE_{\nu}\int_{T_{\rm min}}^{T_{\rm max}(E_{\nu})}\left(\frac{d\sigma}
 {dT}\right)_{\rm SM}^{\rm coh} dT ,
\end{equation}

\noindent where $M_{\text{detector}}$ is the mass of the 
detector under study, $\phi_{0}$ is the total neutrino flux,
$t$ is the data taking time period, $\lambda(E_{\nu})$ is the neutrino
spectrum, $E_{\nu}$ is the neutrino energy, and $T$ is the nucleus
recoil energy.  The maximum recoil energy is related to the neutrino
energy and the nucleus mass through the relation $T_{\rm
  max}(E_{\nu})=2E_{\nu}^{2}/(M+2E_{\nu})$.
  
\noindent
In our analysis, in order to forecast the sensitivity of the CENNS experiments,
we will use two different approaches: we will perform a $\chi^2$
analysis of each proposal, considering that the future experiment will
measure the number of events predicted by the Standard Model. To
compute this values we will use the predicted value for the weak
mixing angle at zero momentum transfer ($\sin^2\theta_W =
0.2386$). With this value as the test experimental value, we will
perform a fit considering different values of the systematic
uncertainties, plus the extreme benchmark case of only statistical
error. A second approach, also used in the present article, will be
the computation of the $\chi^2$ function considering the predicted
statistical error and the systematics coming from the reactor neutrino
spectrum~\cite{Huber:2004xh}, this method has been previously used for the case of
ENES experiments~\cite{Canas:2016vxp}.  For
the reactor neutrino spectrum we will use the expansion discussed in
Ref.~\cite{Mention:2011rk}, while for energies below 2~MeV
the computations reported in Ref.~\cite{Kopeikin:1997ve} were considered. 
In each case we assumed as a benchmark one year of data taking.

\noindent
As already mentioned above, in our first approach we will consider
an analysis based on the function 
\begin{equation}
  \label{Eq:chi}
  \chi^2 = \frac { (N^{\rm SM}_{\rm events}-N^{th})^2 } {\sigma_{stat}^2 + \sigma_{syst}^2} ,
\end{equation}
where the theoretical prediction for the number of events $N^{th}$
will depend on the value of the weak mixing angle and we will
considered different values for the future systematic error
$\sigma_{syst} = p N^{th}/100$, where $p$ will be the percentage of
systematic uncertainty.
For our second approach, we will consider the
current level of uncertainty in the reactor antineutrino spectrum as
an input.

\noindent
We have computed the expected number of events taking into account the
experimental details of each proposal, summarized in
Table~\ref{Tab:01}.  For the RED100 proposal~\cite{Akimov:2012aya} we
consider a $100$~kg target of Xe, a material that is currently of
great interest for coherent scattering~\cite{Baxter:2017ozv} and that
has reached a low energy threshold in different
tests~\cite{Santos:2011ju}. A $500$~eV threshold is expected in the
case of the RED100 experiment. New analyses in this direction are
encouraging and it is expected that the detector will perform even
better~\cite{Akimov:2017gxm}; however, for our analysis we will
restrict to this more conservative estimate. The RED100 experiment
will be located at the Kalinin power plant. In the case of CONNIE, we
consider the most optimistic case of a $1$~kg Si detector, with a
$28$~eV threshold, located at $30$~m from the Angra-2 reactor. As for
the MINER proposal, we perform our computations considering a detector
that will be made of $^{72}$Ge and $^{28}$Si. The proportion between
these two materials is of $2:1$ and the threshold energy is expected
to reach $10$~eV. The antineutrino source in this case will be a
non-comercial TRIGA-type pool reactor that delivers mainly $^{235}$U
antineutrinos~\cite{Dutta:2015vwa}.  We will consider an event rate of
$5$~kg~$^{-1}$~day$^{-1}$~\cite{Agnolet:2016zir} and, as in the case
of all other proposals, one year of data taking.  For the case of
TEXONO, we have considered their proposed High-purity Germanium
detectors as a target with the threshold energy $T_{thres}\sim
100$~eV~\cite{Wong:2008vk,Soma:2014zgm} exposed to an antineutrino
flux coming from the Kuo-Sheng nuclear power plant. Finally, in the
case of the CONUS proposal we follow~\cite{Lindner:2016wff}, where
a detector of up to $100$~kg of germanium is considered, with a recoil
energy threshold as low as $100$~eV.

\begin{table}
  \begin{tabular}{l c c c c c c c c} \hline \hline
& $T_{thres}$ & Baseline & $Z/N$  & Det. Tec. & Fid. Mass  \\ \hline \hline
CONNIE~\cite{Aguilar-Arevalo:2016qen,Aguilar-Arevalo:2016khx} 
& $28$~eV &30~m& $1.0$ & CCD (Si) & $1$~kg\\ 
  RED100~\cite{Akimov:2012aya}  
&  500~eV  &19 m & $0.70$  &Lq.Xe& $100$~kg\\
MINER~\cite{Agnolet:2016zir}    & 
10~eV  & $1$ m & $0.81 $   &    $^{72}$Ge:$^{28}$Si (2:1) & $30$~kg \\
TEXONO~\cite{Soma:2014zgm}   & 
100~eV  &28 m & $0.79$  & HPGe & $1$ kg\\ 
CONUS~\cite{Lindner:2016wff}  &    
 100~eV & 10~m & $0.79$  & HPGe & $100$~kg \\ \hline \hline 
\end{tabular}\caption{\label{Tab:01} List of some experimental proposals to detect CENNS with reactor antineutrinos. }
\end{table}

\begin{figure}[ht] 
\begin{center}
\includegraphics[width=0.328\textwidth]{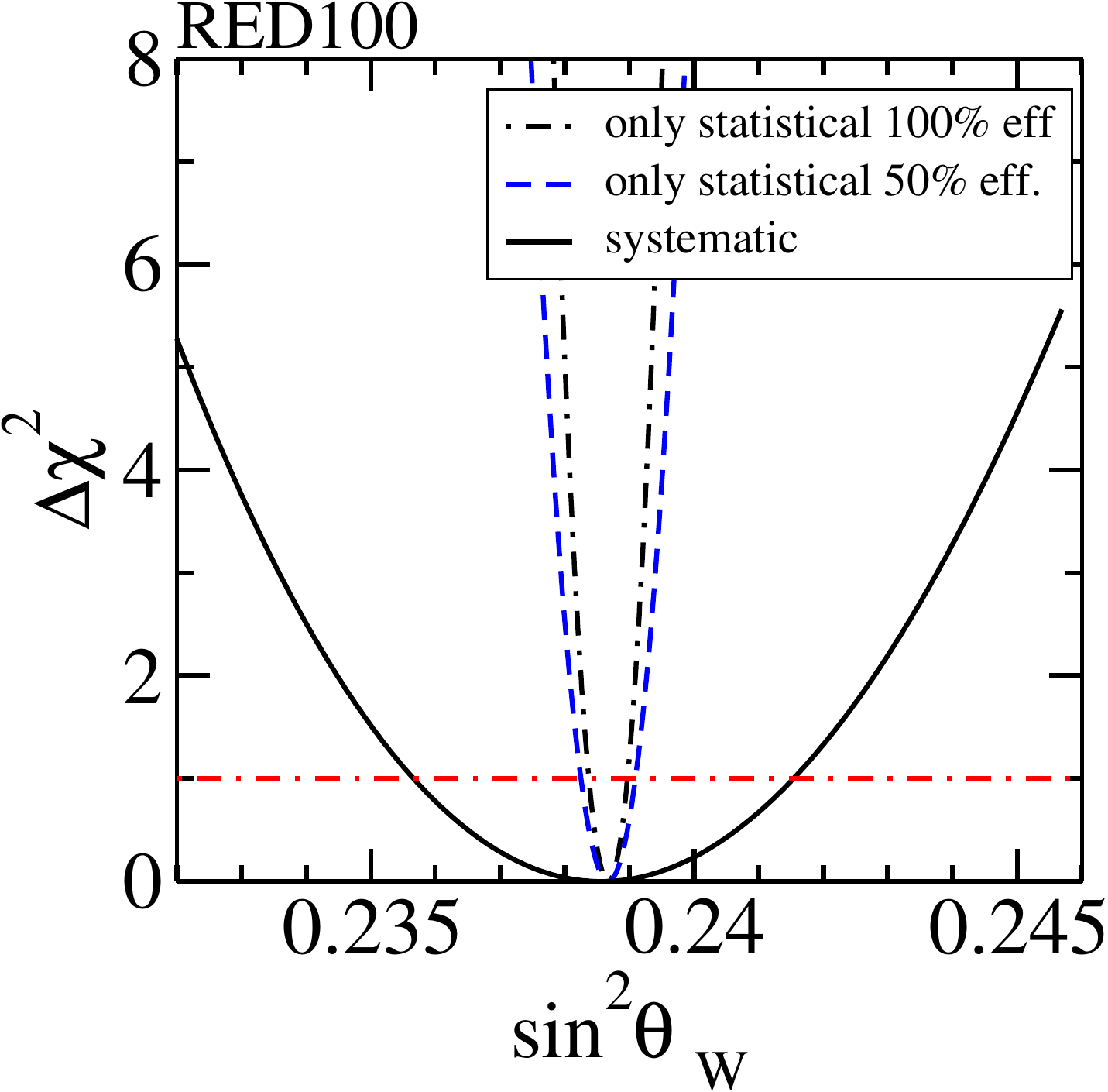}
\includegraphics[width=0.328\textwidth]{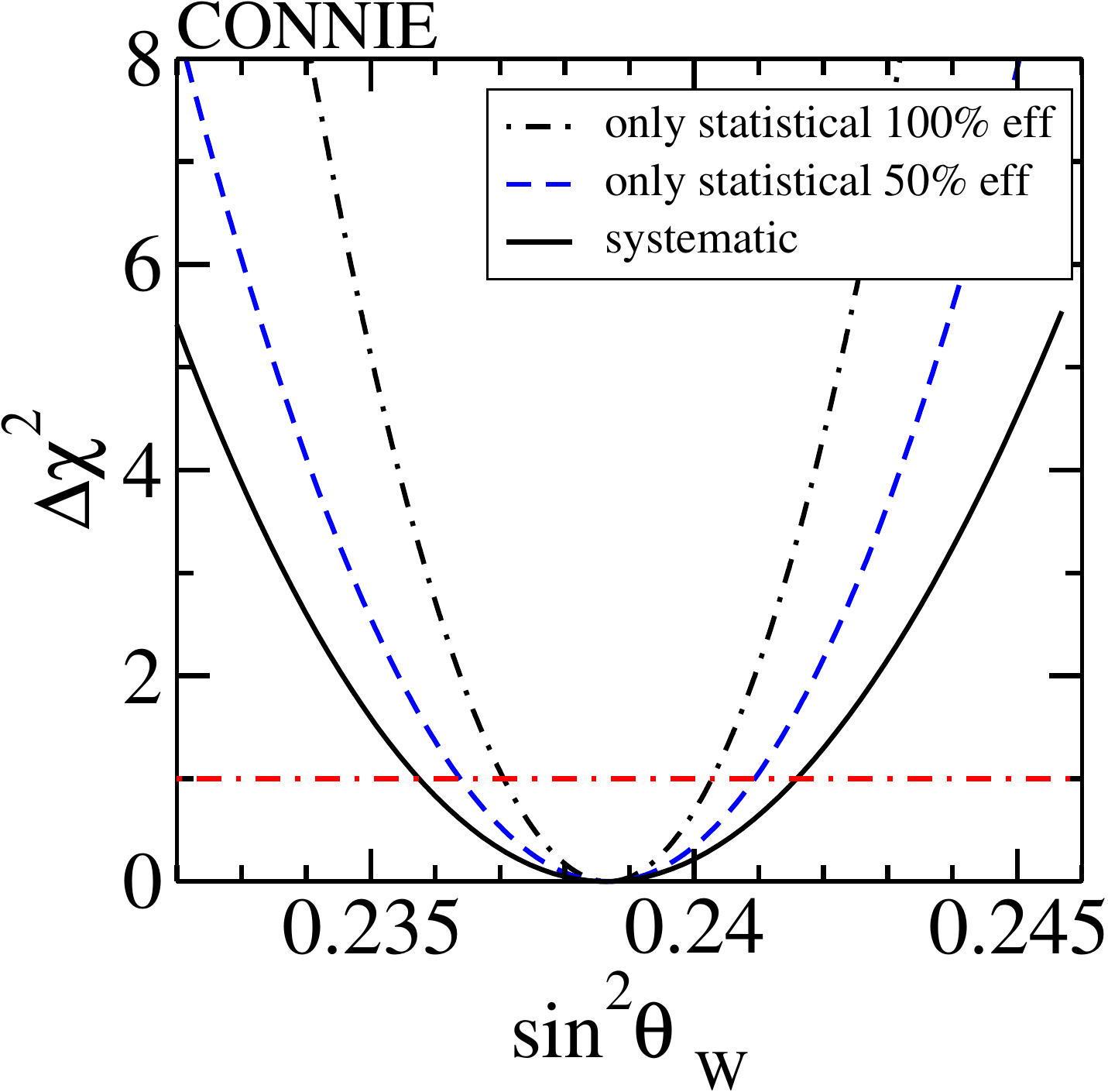}
\includegraphics[width=0.328\textwidth]{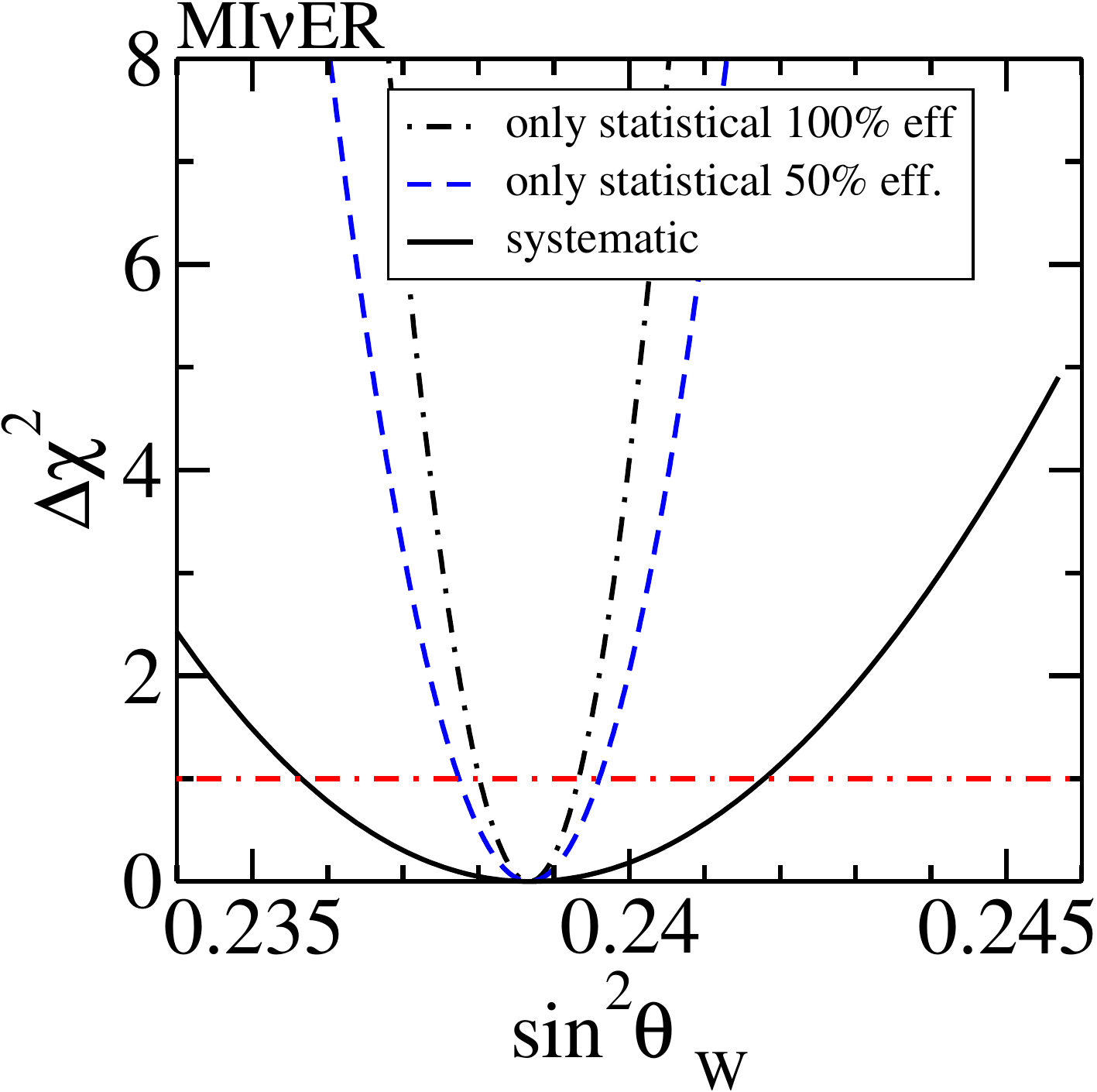}
\end{center}
\caption{\label{fig:wma} Expected sensitivity of the RED100(left), CONNIE(center) and MINER(right)  detectors to the weak mixing angle.
 The dotted-dashed line (black) and the dashed line (blue) are the curves considering only statistical errors, 
 and they correspond to a 100\% and 50\% efficiency, respectively. 
 The case including systematic errors from the reactor neutrino spectra with a 100\% efficiency is shown by the solid (black) line.}
\end{figure}

\section{Weak Mixing Angle sensitivity}
With the information given above, we have computed the expected
sensitivity to the weak mixing angle, $\sin^2 \theta_W$. We have
assumed that the future experimental setups will measure exactly the
Standard Model prediction and computed the corresponding fit as
mentioned in Eq.~(\ref{Eq:chi}) for three different cases:  (i) when the
experiment is capable of an optimal efficiency (100 \%), (ii) when it
reaches an efficiency of $50$~\%, and (iii) in the case when we include the
current systematic uncertainty corresponding to the theoretical
antineutrino flux, with a statistical error corresponding to a
$100$~\% efficiency. We can see the results of this analysis in
Fig.~(\ref{fig:wma}), where we show the cases of
CONNIE~\cite{Aguilar-Arevalo:2016qen,Aguilar-Arevalo:2016khx,Chavez-Estrada:2017gni},
MINER~\cite{Agnolet:2016zir} and the RED100~\cite{Akimov:2012aya}
proposals.
For the value of the weak mixing angle, we have considered the extrapolation
to the low energy regime: 
\begin{equation}
\sin^2\theta_W(0)_{\overline{\text{MS}}} 
= \kappa(0)_{\overline{\text{MS}}}  \sin^2\theta_W(M_Z)_{\overline{\text{MS}}} 
\end{equation}
with $\kappa(0) = 1.03232$~\cite{Kumar:2013yoa}.

\noindent
From Fig.~(\ref{fig:wma}) we can notice that the perspectives for a
precise measurement of the weak mixing angle are promising, and that
they are dominated by the systematic error from the reactor
spectrum. However, it is expected that this error will be reduced,
thanks to the progress in the current knowledge of the reactor spectrum from
 its direct measurement at IBD experiments. We can also notice that for the case of the CONNIE
collaboration, it will be necessary to have a higher mass detector in order to
reduce the statistical error. This is due to the fact that the
detector has very low mass and the target material is also lighter.  We
show in Table~\ref{Tab:02}, the corresponding $1\sigma$ error for
$\sin^2\theta_W$ for the three different configurations under
discussion. We have also included for comparison the results for CONUS
and TEXONO. We can see that the results can be competitive, especially
if systematical errors can be reduced.

\begin{figure}[ht] 
\begin{center}
\includegraphics[width=0.8\textwidth]{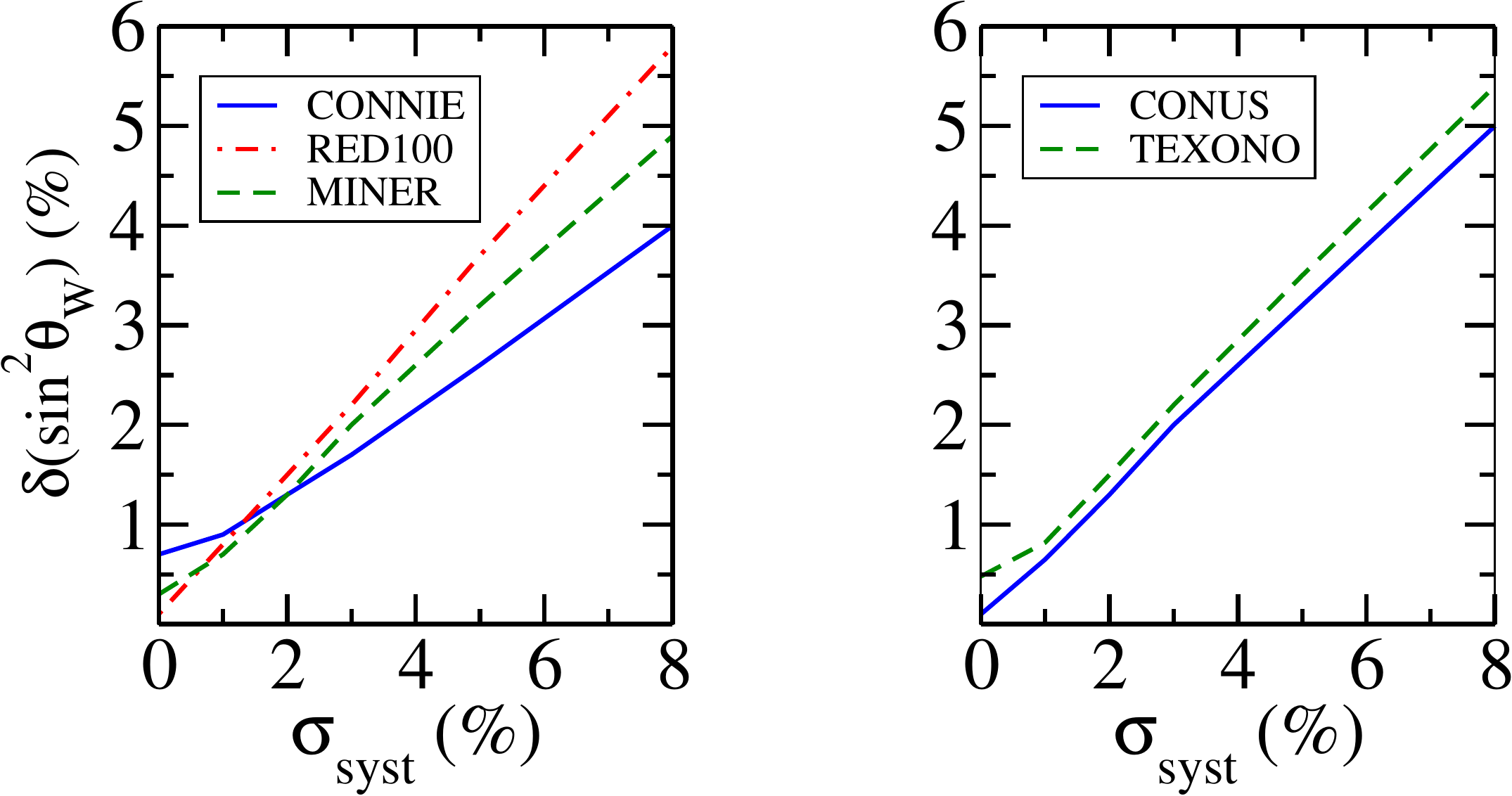}
\end{center}
\caption{\label{fig:percent} Expected sensitivity to $\sin^2\theta_W$
  (in percent) for the different proposals under consideration,
  depending on the systematic uncertainty to be achieved, in
  percent. In the left panel is shown the expected error on the weak mixing angle 
  for the experiments under study in Fig.~(\ref{fig:wma}). 
  In the right panel are shown TEXONO and CONUS, two proposals that use the same
  nucleus as a target and, therefore, have a similar dependence.  }
\end{figure}

\noindent
In order to have a better idea of the dependence of the sensitivity on
the systematics, we have plotted in Fig.~(\ref{fig:percent}) the
expected error on the weak mixing angle, depending on the systematic
error that each particular experiment can reach. In this case, we have
also included the result for the Texono and the Conus proposals. From
this figure, it is possible to see that CONNIE is slightly less
affected by the systematics than other experiments. Being an
experiment where the proportion of protons to neutrons is higher, this
result seems natural, while among Texono and CONUS, the
dependence is very similar, since they use the same
target material.

\begin{table}
  \begin{tabular}{l c c c c c c  } \hline \hline 
    &  $50$ \%  & eff.  & $100$ \% & eff.  & including & systematics \\ 
  experiment  &  $\delta_{\sin^2\theta_W}$ & \% &   $\delta_{\sin^2\theta_W}$ & \%
&   $\delta_{\sin^2\theta_W}$ & \%  \\
  \hline \hline
%& 100\% + stat. err. & & &realistic\\ 
TEXONO        & 0.0015 & 0.6 & 0.0011 & 0.5 &  0.0028 & 1.2\\ 
RED100        & 0.0004 & 0.2 & 0.0003 & 0.1 &  0.0031 & 1.3 \\
MINER         & 0.0010 & 0.4 & 0.0007 & 0.3 &  0.003  & 1.3 \\
CONNIE        & 0.0023 & 1.0 & 0.0017 & 0.7 &  0.003  & 1.3 \\ 
CONUS     & 0.0003 & 0.1 & 0.0002 & 0.1 &  0.0023 & 1.0 \\ \hline \hline 
\end{tabular}\caption{\label{Tab:02} Expected sensitivity to the weak mixing angle. For each experiment we quote the 
$1\sigma$ expected sensitivity in the case of a 50 \%
    (100\%)efficiency of the experiment and for the case of a
    systematic error equal to that of the current reactor spectrum
    uncertainty. The results are shown in terms of
    $\delta(\sin^2\theta_W)$ as well as in percent.}
\end{table}

\section{Discussion and conclusions}
\begin{figure}[ht] 
\begin{center}
\includegraphics[width=0.75\textwidth]{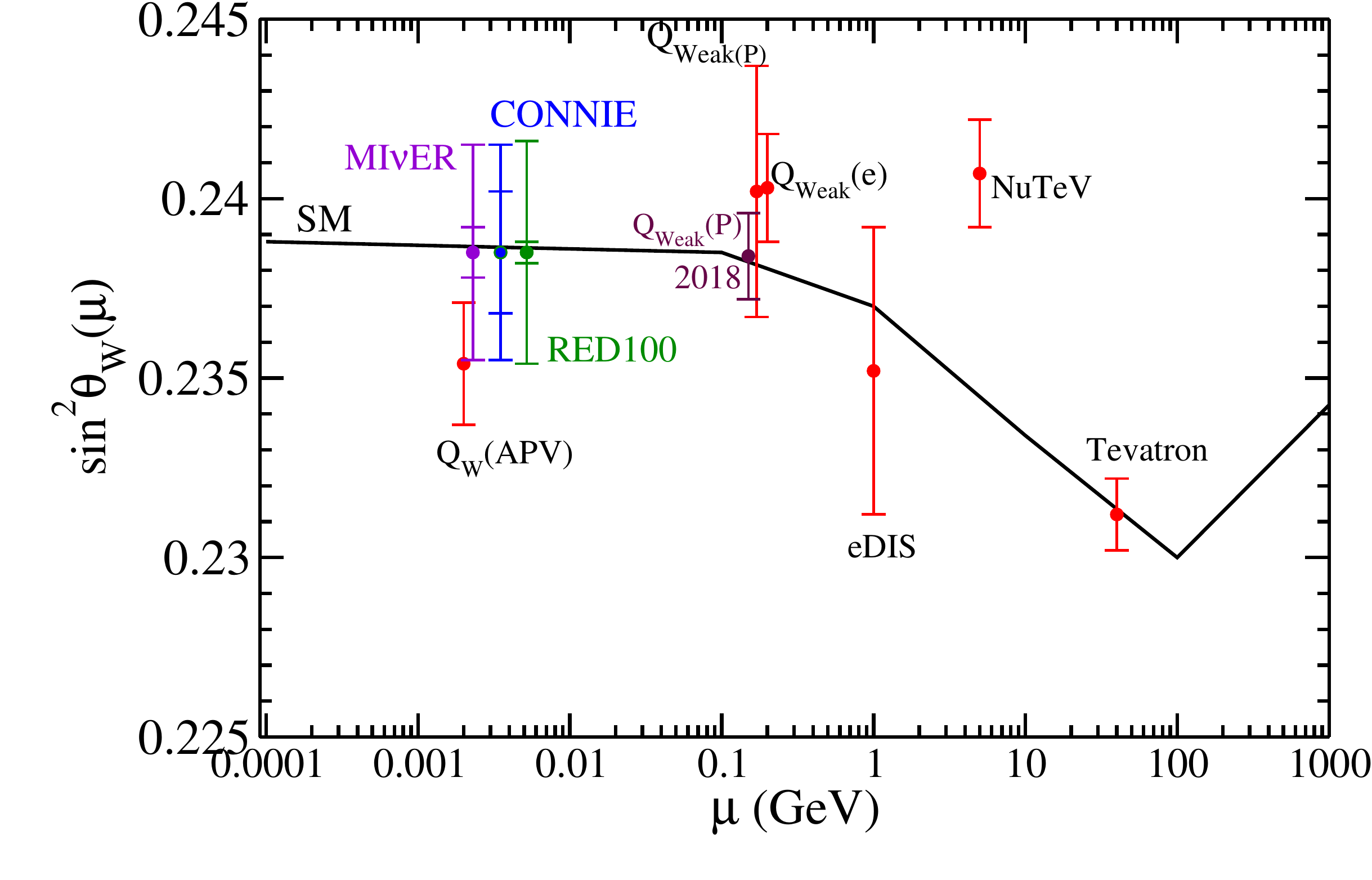}
\end{center}
\caption{\label{fig:PDG} Expected sensitivity of CENNS experiments to the weak mixing angle compared with the SM prediction~\cite{Erler:2004in,Erler:2017knj}, 
in the $\overline{MS}$ renormalization scheme. 
Electron weak charge  $Q_{W(e)}$ comes from Moller scattering~\cite{Anthony:2005pm}, and both
the former~\cite{Androic:2013rhu} and recent~\cite{Androic:2018kni} measurements of the proton weak charge $Q_{Weak(P)}$ are also shown.}
\end{figure}
The weak mixing angle is one of the fundamental parameters of the
Standard Model and it has been measured with great accuracy at the
$Z$-pole~\cite{Patrignani:2016xqp}. At very low momentum transfer
there are also measurements of this important quantity, although the
precision is lower. The main results in this energy window come from
the measurement of the weak charge, such as in the recent measurement
by Qweak~\cite{Androic:2018kni}, and from atomic parity violation
experiments~\cite{Patrignani:2016xqp}, a measurement that will be
improved by the P2~\cite{Becker:2018ggl}, SoLID~\cite{Souder:2016xcn}
 and Moller~\cite{Benesch:2014bas}
experiments.  Both measurements
are extracted from the weak charge in protons or electrons.
The measurement of the weak mixing angle at the low energies in
neutrino scattering processes has plenty of room for
improvement~\cite{Canas:2016vxp} and the CENNS experiments have the
potential to obtain a competitive accuracy, provided that systematic
errors can be reduced.

\noindent
In this work we have computed the expected sensitivity for different
CENNS proposals and we have shown the viability of such a measurement with a
reasonable accuracy. Moreover, if the systematic errors can be
reduced, the measurement of the weak mixing angle from CENNS experiments can be even better
than the one coming from electron weak charge. We show this potential in
Fig~(\ref{fig:PDG}) the result of Table~\ref{Tab:02} is presented  in
a graphical representation comparing the future measurement of the
weak mixing angle in CENNS with current measurements. We can see that
the CENNS experiments can really give a good measurement of this
observable through a different and new channel. 

\acknowledgments { This work was supported by CONACYT-Mexico, SNI (Sistema Nacional de Investigadores), and PAPIIT project IN113916. A. Parada was
  supported by Universidad Santiago de Cali (USC) under grant 935-621118-3.
  }
  
%  \bibliography{merged} 

%
\end{document}